\documentclass{article}

\usepackage[preprint]{neurips_2022}

\usepackage[utf8]{inputenc} 
\usepackage[T1]{fontenc}    
\usepackage{hyperref}       
\usepackage{url}            
\usepackage{booktabs}       
\usepackage{amsfonts}       
\usepackage{nicefrac}       
\usepackage{microtype}      
\usepackage{xcolor}         
\usepackage{graphicx}

\title{SubZero: Subspace Zero-Shot MRI Reconstruction}

\author{%
    Heng Yu\\
The Robotics Institute \\
Carnegie Mellon University \\
  Pittsburgh, PA \\
  \And
  Yamin Arefeen \\
  Department of Electrical Engineering and Computer Science \\
  Massachusetts Institute of Technology \\
  Cambridge, MA \\
  \AND
  Berkin Bilgic \thanks{Also with Department of Radiology, Harvard Medical School, Boston, MA} \\
  Athinoula A. Martinos Center for Biomedical Imaging \\
  Massachusetts General Hospital \\
  Charlestown, MA \\
}

\begin{document}

\maketitle

\begin{abstract}
Recently introduced zero-shot self-supervised learning (ZS-SSL) has shown potential in accelerated MRI in a scan-specific scenario, which enabled high quality reconstructions without access to a large training dataset. ZS-SSL has been further combined with the subspace model to accelerate 2D $T_2$-shuffling acquisitions. In this work, we propose a parallel network framework and introduce attention mechanism to improve subspace based zero-shot self-supervised learning and enable higher acceleration factors. We name our method SubZero and demonstrate that it can achieve improved performance compared with current methods in $T_1$ and $T_2$ mapping acquisitions.
\end{abstract}

\section{Introduction}

Deep learning methods have proven to be powerful in MRI reconstruction. However, they often require large training data which is not always available, and differences in anatomy/pathology between the training and test data may lead to poor generalization. Yaman et al.\ [1] proposed a zero-shot self-supervised learning method (ZS-SSL) for scan-specific MRI reconstruction and this has been further extended by incorporating $T_2$-shuffling forward model (ZSSSSub). TheZSSSSub method outperforms standard $T_2$-shuffling and ZS-SSL by modifying the forward model and keeping the original network structure of ZS-SSL. We demonstrate that a tailored network structure may achieve better results beyond those obtained with  ZSSSSub. In this abstract, we propose SubZero which introduces a parallel network framework inspired by previous work\ [2] and leveraging attention mechanism. In vivo reconstruction experiments suggest that our SubZero outperforms current works at high acceleration rates of $R=8-10$.

\section{Methods}
\subsection{Zero-shot self-supervised learning (ZS-SSL)}

ZS-SSL split the undersampled kspace data $\Omega$ into three subsets: $\Theta$ for the data consistency units, $\Lambda$ for k-space training loss in self-supervision and $\Gamma$ for automated early stopping. The reconstruction process can be formulated into two sub-problems\ [3]:

$$\textbf{z}^{(i-1)}=\arg\min_{\textbf{z}}\mu\|\textbf{x}^{(i-1)}-\textbf{z}\|^2_2+R(\textbf{z})$$
$$\textbf{x}^{(i)}=\arg\min_{\textbf{x}}\|\textbf{y}_\Omega-\textbf{E}_\Omega \textbf{x}\|^2_2+\mu\|\textbf{x}-\textbf{z}^{(i-1)}\|^2_2$$




\subsection{Subspace model based ZSSS (ZSSSSub)}

Subspace method first assembles the signal evolution into a dictionary and then approximates the low rank subspace $\Phi$ using singular vector decomposition (SVD). The $T_1$ or $T_2$ weighted images $x$ to be reconstructed can be projected onto $\Phi$ and yield coefficients $\alpha$, where $x\in C^{M\times N\times T}$, $\Phi \in C^{T\times B}$, $\alpha \in C^{M\times N\times B}$, and $M$, $N$ are the number of voxels in readout and phase encoding directions, $T$ is echo number and $B$ is basis number. The subspace forward model can be written as $y=RFS\Phi\alpha$, where $y\in C^{M\times N\times C\times T}$ is the acquired kspace data, $R$ is the undersampling mask, $S$ is the coil sensitivity map and $F$ is the Fourier operator. ZSSSSub incorporates low rank subspace model into ZSSS and can be formulated as\ [4]:

$$\textbf{z}^{(i-1)}=\arg\min_{\textbf{z}}\mu\|\alpha^{(i-1)}-\textbf{z}\|^2_2+R(\textbf{z})$$
$$\alpha^{(i)}=\arg\min_{\alpha}\|\textbf{y}_\Omega-\textbf{E}_\Omega \Phi \alpha\|^2_2+\mu\|\alpha-\textbf{z}^{(i-1)}\|^2_2$$

\subsection{Proposed method (SubZero)}
\begin{figure}
  \centering
  \includegraphics[width=0.95\textwidth]{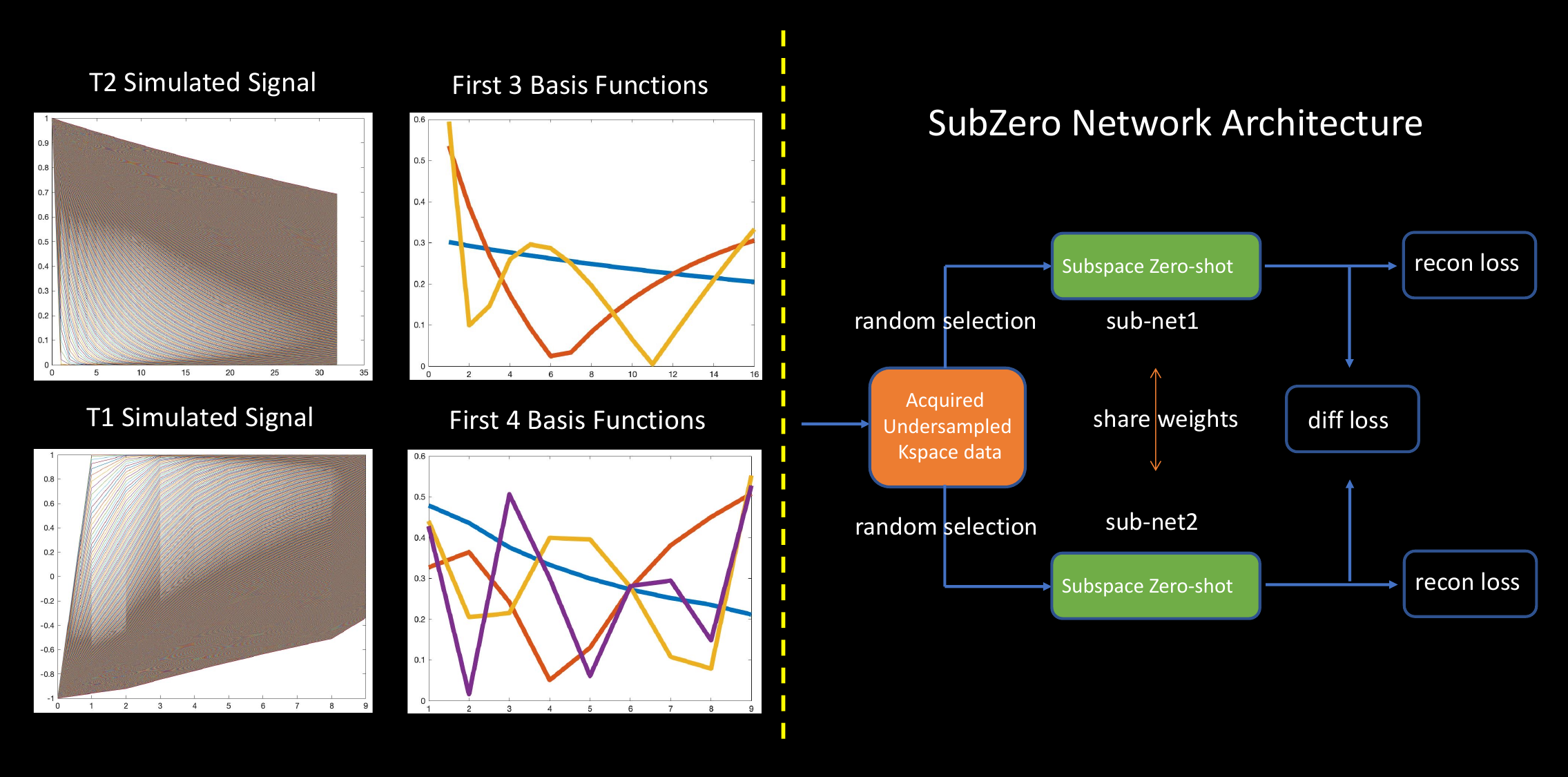}
  \caption{Left: $T_2$/$T_1$ simulated signal dictionary and first few basis of SVD. Right: Overall framework of our proposed SubZero. SubZero includes two sub-networks and takes in multi-echo data with shifted under-sampled mask.}
  \label{fig1}
\end{figure}

\begin{figure}
  \centering
  \includegraphics[width=0.5\textwidth]{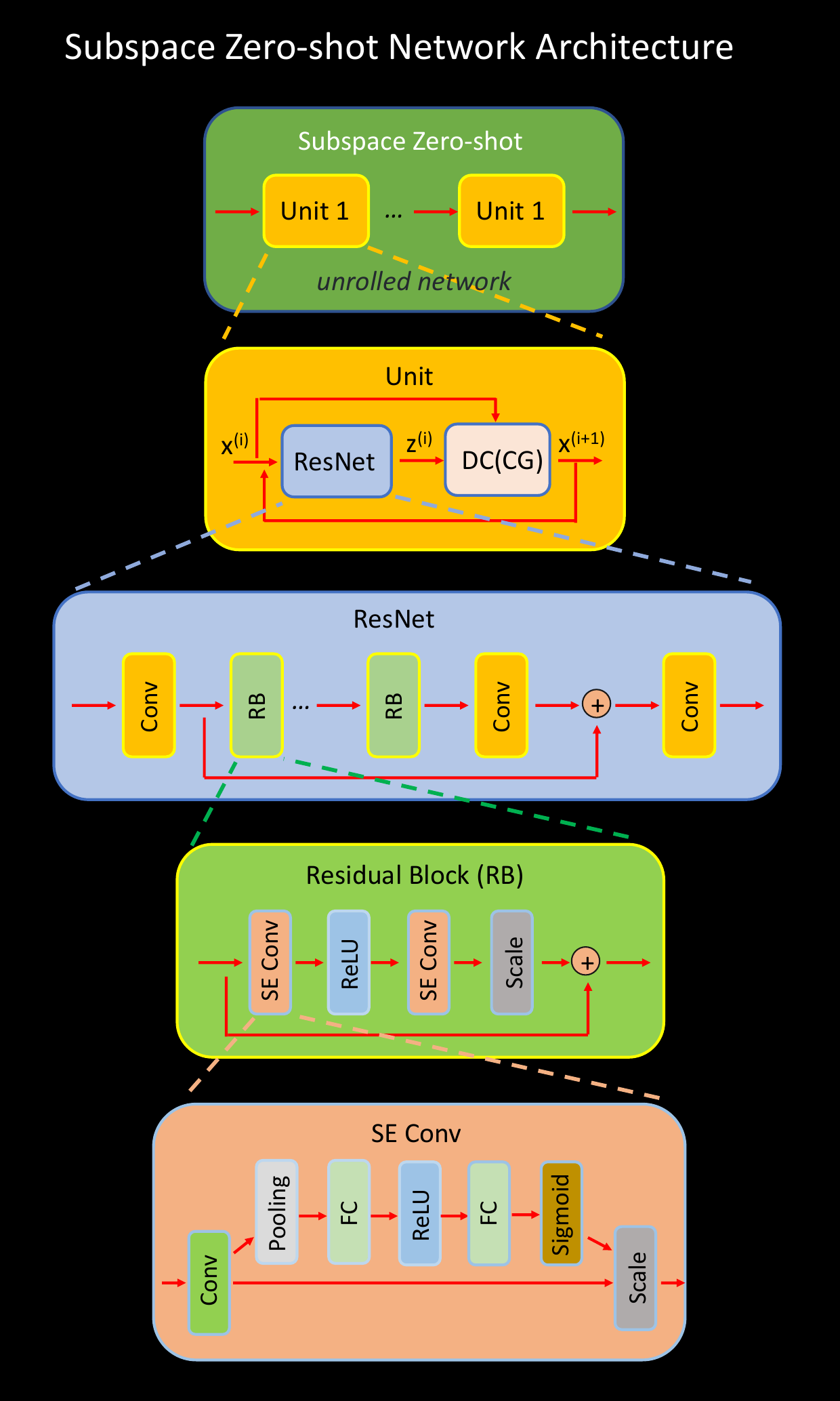}
  \caption{Network structure of each sub-network. We introduce SE convolution which is a kind of attention mechanism based on ZSSSSub method.}
  \label{fig2}
\end{figure}

Our SubZero has two parallel sub-networks as shown in Figure~\ref{fig1}. The two sub-networks have the same structure and share weights. Each sub-network can be seen as a ZSSSSub module as shown in Figure~\ref{fig2} and we replace the traditional convolution with Squeeze-and-Excitation (SE) convolution proposed in previous work\ [5] which introduces attention mechanism to help better learn the proper features. SE convolution performs a squeeze operation to get a single value for each channel of the input feature map and then learns per-channel weights using an excitation operation. The final output of the SE convolution is obtained by rescaling the feature map with these per-channel weights. The automated early stopping mask $\Gamma$ is the same for two sub-networks and $\Theta$, $\Lambda$ are randomly selected from $\Omega \backslash \Gamma$ for each sub-network. The total loss consists of reconstruction loss and difference loss as follows:

$$L=L^{recon}_1+L^{recon}_2+L^{diff}$$
$L^{recon}_i$ ($i \in {1, 2}$ from individual sub-networks) is the same as ZSSSSub and $L^{diff}$ forces the reconstruction results from two sub-networks to be consistent in $\Omega \backslash (\Theta \cup \Lambda)$. Note $\Omega \backslash (\Theta \cup \Lambda)$ as $\Delta$, $X^{(1)}$ and $X^{(2)}$ as the reconstruction results from two sub-networks, we have:
$$L^{diff}=MSE(X^{(1)}_\Delta, X^{(2)}_\Delta)$$
We also use a novel approach to mask generation when dividing $\Theta$ and $\Lambda$ during the training process. Given a pre-defined ratio $r$, we randomly generate the division mask among all the echoes since we use $T_1$ or $T_2$ series data to provide complementary information while  keeping the whole mask ratio as $r$. This online data augmentation introduces more diversity during training. We also shift the under-sampling mask $\Omega$ among echoes to achieve better performance.

\section{Experiments and Results}
\begin{figure}
  \centering
  \includegraphics[width=0.95\textwidth]{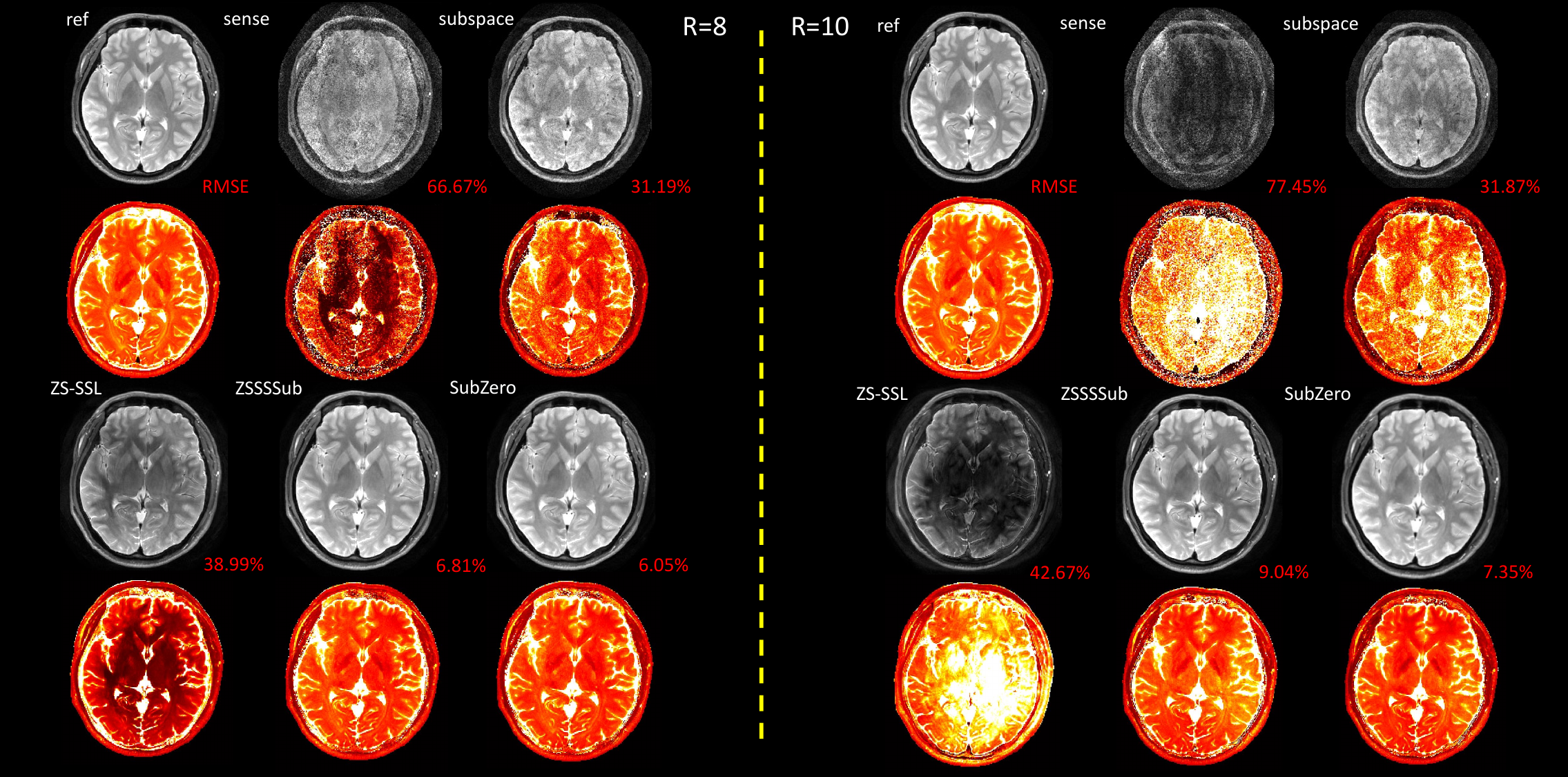}
  \caption{Reconstruction examples of $T_2$-weighted images (B = 3) and T2 mapping examples using corresponding reconstruction images at acceleration rate R = 8 and 10. ref: ground-truth. sense: sense method. subspace: traditional subspace method. ZSSS: zero-shot self-supervised learning method. ZSSSSub: Subspace model based ZSSS. SubZero: Proposed method. Here we show the second echo and the red number indicates the average RMSE of all 16 echoes.}
  \label{fig3}
\end{figure}

\begin{figure}
  \centering
  \includegraphics[width=0.95\textwidth]{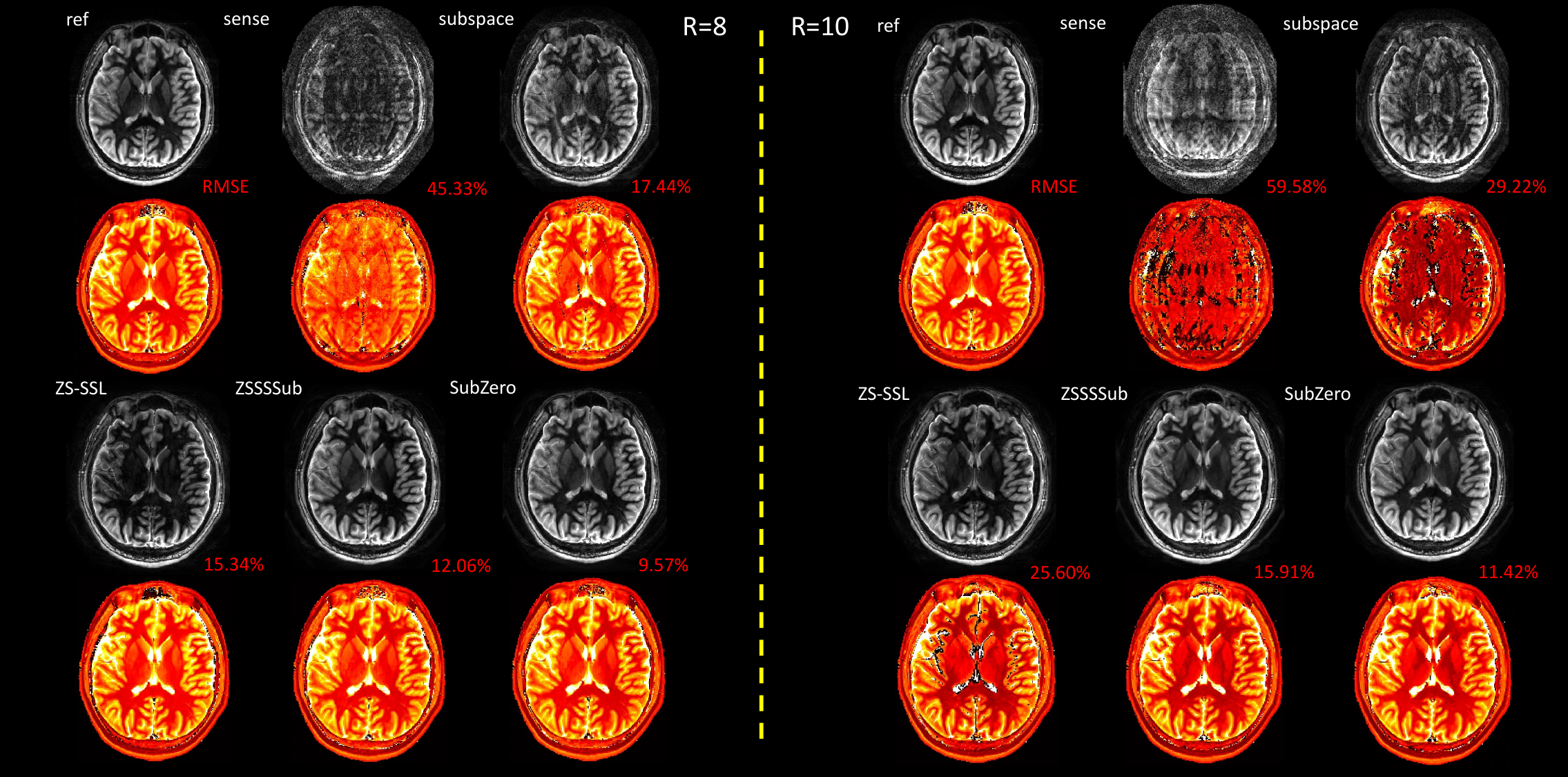}
  \caption{Reconstruction examples of $T_1$-weighted images (B = 4) and T1 mapping examples using corresponding reconstruction images at acceleration rate R = 8 and 10. Here we show the third echo and the red number indicates the average RMSE of all 9 echoes.}
  \label{fig4}
\end{figure}

\begin{figure}
  \centering
  \includegraphics[width=0.95\textwidth]{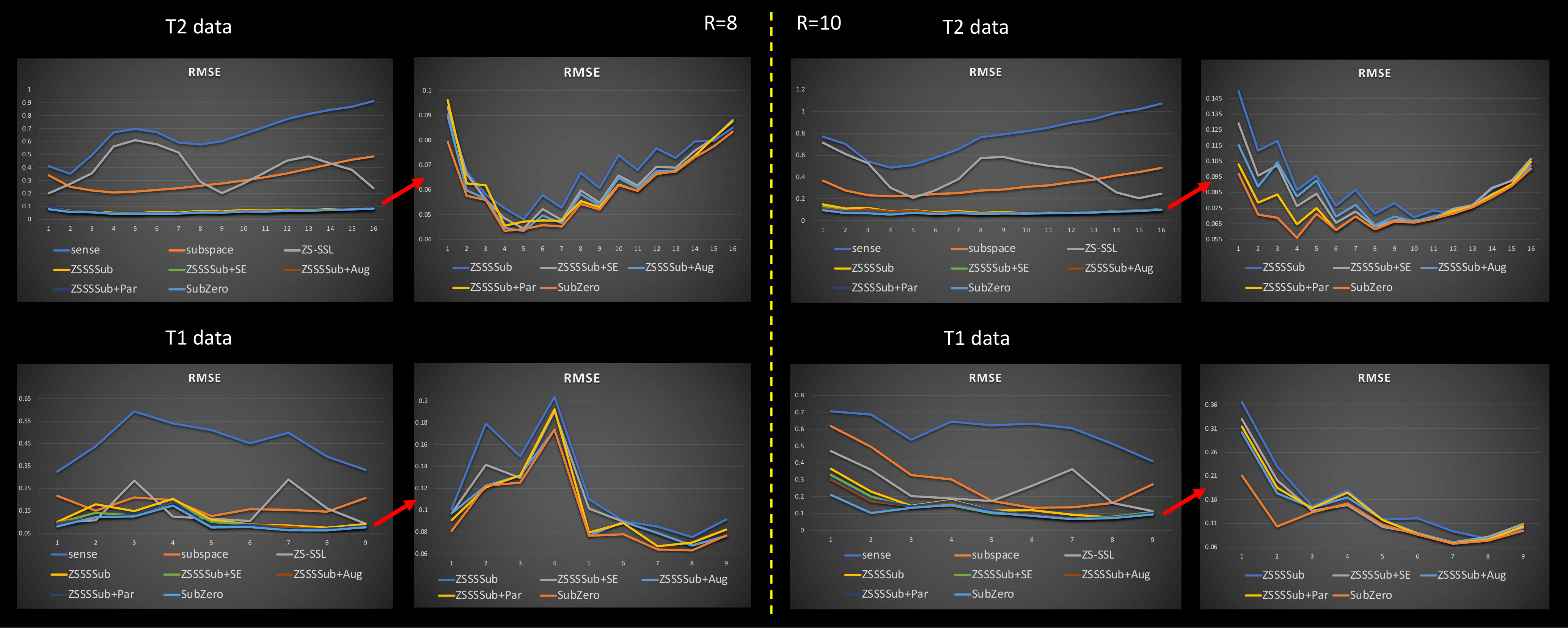}
  \caption{RMSE of each echo using different methods. ZSSSSub+SE: Incorporating SE convolution into ZSSSSub. ZSSSSub+Aug: Using proposed online data augmentation method in ZSSSSub. ZSSSSub+Par: Using parallel framework with each sub-network being ZSSSSub (no SE convolution and proposed online data augmentation method).}
  \label{fig5}
\end{figure}

We conduct experiments using $T_2$ and $T_1$-weighted images. $T_2$-weighted multi-echo spin echo data includes 16 echoes with 12 channels and TE=11.5-368 ms ($\delta$TE=11.5 ms). The full FOV is $256 \times 208$ and we take $B=3$ for SVD of simulated signal dictionary. $T_1$-weighted inversion recovery turbo spin echo data includes 9 inversion times (TIs) with 32 channels and TI=100-2000 ms. The full FOV is $256 \times 192$ and we take $B=4$ for SVD of simulated signal dictionary. We use acceleration rates of R 8 and 10 along the phase encoding direction for both $T_2$ and $T_1$-weighted image data. For SubZero model, we set  unrolled block number as 5, resnet block number 10, and iteration of conjugate gradient (CG) as 5. From Figure~\ref{fig3} and Figure~\ref{fig4} we can see that our SubZero outperforms other models at varying acceleration rates. We also perform ablation experiments to show each of our proposed modules can help improve the performance as shown in Figure~\ref{fig5}. Incorporating SE convolution into ZSSSSub, using proposed online data augmentation method in ZSSSSub and using parallel framework with each sub-network being ZSSSSub are all proven effective to improve the performance of ZSSSSub, each of them can achieve up to 2-fold RMSE gain.
Code/data are available at \url{https://github.com/Heng14/SubZero}.



\section{Discussion and Conclusion}

In this abstract, we propose SubZero method which incorporates subspace model into a zero-shot self-supervised learning framework by using a parallel network and attention mechanism along with a new online data augmentation method. We conduct experiments on multi-echo $T_2$ and multi-inversion $T_1$-weighted images and perform ablation studies of each proposed module. Experimental results demonstrate reductions of up to 4-fold RMSE over ZS-SSL, 3-fold over ZSSSSub using SubZero, especially at higher acceleration rates.

\begin{ack}

This work was supported by research grants NIH R01EB028797, R01EB032378, U01EB025162, P41EB030006, U01EB026996, R03EB031175 and the NVidia Corporation for computing support.


\end{ack}

\section*{References}


{
\small




[1] Yaman, Burhaneddin, Seyed Amir Hossein Hosseini, and Mehmet Akcakaya. "Zero-Shot Self-Supervised Learning for MRI Reconstruction." International Conference on Learning Representations. 2021.

[2] Hu, Chen, et al. "Self-supervised learning for mri reconstruction with a parallel network training framework." International Conference on Medical Image Computing and Computer-Assisted Intervention. Springer, Cham, 2021.

[3] Yaman, Burhaneddin, et al. "Self‐supervised learning of physics‐guided reconstruction neural networks without fully sampled reference data." Magnetic resonance in medicine 84.6 (2020): 3172-3191.

[4] Zhang, Molin, et al. “Zero-Shot Self-Supervised Learning for 2D T2-shuffling MRI Reconstruction.” Proceedings of the International Society of magnetic Resonance in Medicine (ISMRM), 2022.

[5] Hu, Jie, Li Shen, and Gang Sun. "Squeeze-and-excitation networks." Proceedings of the IEEE conference on computer vision and pattern recognition. 2018.

}

\end{document}